\documentclass[twocolumn,runningheads]{svjour2}
\smartqed  
\usepackage{graphicx}
\journalname{Astrophysics and Space Science}
\begin{document}

\title{Identification of high energy gamma-ray sources and source populations in the era of deep all-sky coverage}

\titlerunning{Identification of new source populations in the GLAST era}        

\author{Olaf Reimer \and Diego F. Torres}

\authorrunning{Olaf Reimer \& Diego F. Torres} 

\institute{O. Reimer \at
           W.W. Hansen Experimental Physics Laboratory \& \\
           Kavli Institute for Particle Astrophysics and Cosmology\\
		   Stanford University \\
           Stanford, CA 94305-4085, USA\\
           Tel.: +1-650-724-6819, Fax: +1-650-725-2463\\
             \email{olr@stanford.edu }         
           \and
		   D. Torres \at
		   Instituci\'o de Recerca i Estudis Avan\c{c}ats (ICREA) \& \\
           Institut de Ci\`encies de l'Espai (IEEC-CSIC)\\
           Facultat de Ciencies, Universitat Aut\`onoma de Barcelona\\
           Torre C5 Parell, 2a planta\\ 
		   08193 Barcelona, Spain\\
		   Tel.: +34-93-581-4352 Fax: +34-93-581-4363  \\
             \email{dtorres@ieec.uab.es} 
}
\date{Received: date / Accepted: date}

\maketitle

\begin{abstract}

A large fraction of the anticipated source detections by the Gamma-ray Large Area Space Telescope 
(GLAST-LAT) will initially be unidentified. We argue that traditional approaches to identify 
individuals and/or populations of gamma ray sources will encounter procedural limitations. 
Those limitations are discussed on the background of source identifications from EGRET observations.
Generally, our ability to classify (faint) source populations in the anticipated GLAST 
dataset with the required degree of statistical confidence will be hampered by sheer source wealth. 
A new paradigm for achieving the classification of gamma ray source populations is discussed. 
\end{abstract}

\section{Problem statement}

The anticipated source wealth from observations carried out by the
satellite-based $\gamma$-ray mission GLAST, potentially yielding the
discovery of thousands of new high-energy sources following
extrapolations from predecessor experiments, will create several
problems for source identification. Catalogs of the most prominent 
candidate sources (Active Galactic Nuclei -AGNs-, and neutron stars/pulsars -PSRs-)
will very likely not be complete to the required low radio and/or X-ray flux levels 
required for counterpart studies (AGN), or do not have the ability to provide
a suitable counterpart at all (radio-quiet PSRs). Predictably, this will leave 
many of the new $\gamma$-ray source detections initially unidentified. 
And even if the pulsar and AGN catalogs were sufficiently deep, they may not 
yield unambiguous source identifications: A complete catalog for the
anticipated numbers of sources, projected using the instrumental
point-spread-function (psf), would generate total sky coverage,
with one or more candidates in every line-of-sight for incident
photons corresponding to their (energy dependent) psf. There would be 
one or more AGN everywhere, and one or more pulsar in every line--of--sight 
at low galactic latitude. This will limit or even prevent unambiguous source
identifications based solely on spatial correlation. In addition, a 
legacy from the EGRET experiment is the indication that we are already 
missing the finite identification of one or more source populations, both 
at low and at high Galactic latitude. Specifically, the identification of 
variable, non-periodic, point-like sources at low galactic latitude, as well
as of non-variable sources at high latitude is still missing \cite{Rei01}, since 
these source populations exhibit characteristics different from the EGRET-detected pulsars or blazars.

In the GLAST era and beyond, if it is the objective to conclusively
identify all individual $\gamma$-ray source detections, we will
predictably fail. The anticipated number of counterparts, their
relative faintness deduced from luminosity functions, the missing
all-sky coverage in the relevant wavebands for deep counterpart
studies, and the expected ambiguities due to source confusion in
densely populated regions of the $\gamma$-ray sky will preclude
reaching this ultimate goal of source identification.
Consequently, we should aim to identify at least all {\it classes}
of sources, and subsequently attempt to gain in-depth
astrophysical knowledge by studying the most interesting or prominent 
representatives among such populations. 
\begin{figure}[ht]
   \centering
   \includegraphics[width=0.48\textwidth]{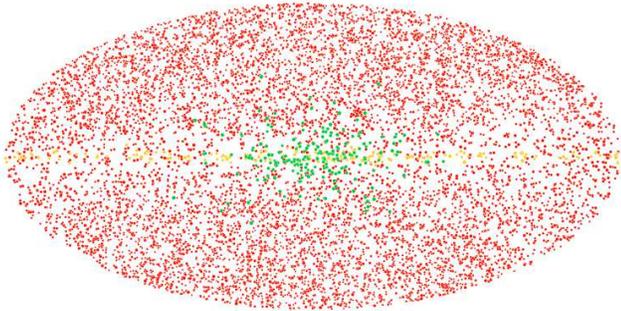} 
   \caption{Synthetic gamma-ray source catalog for GLAST-LAT observations, based on a flux-limited source sample
   according to a realistic diffuse gamma-ray emission model. AGN (red dots) dominate the catalog.  Two additional 
   source populations have been considered: Sources in the Galactic bulge (yellow), and a Galactic halo source population (green). 
   Figure credits: Seth Digel}
   \label{Fig.1}
\end{figure}
The anticipated number of source detections left unidentified will 
preclude individual deep multifrequency studies for every source, in 
the way it led to the identification of e.g. the Geminga pulsar and 
various $\gamma$-ray blazars.  

Suppose that we have a sufficiently complete counterpart catalog,
such that a member of it spatially coincides with most of the GLAST-LAT
sources. Does this imply that we have already identified all
sources? To answer this question consider that we have, instead, a
reasonably complete sky coverage of sources, i.e. GRBs as an
example. An overlay of all error boxes of GRBs reported from BATSE
covers the whole sky. Then, there is at least one GRB spatially
coinciding with any possible counterpart or host. Consequently,
here a spatial correlation analysis lacks identification capability, 
even when it is clear that not all populations of astrophysical objects
are plausible candidates for GRB generation or hosting, nor that
all of them should even be probed. More particularly, we can not claim,
using correlation analysis, that GRBs have appeared more often in
starburst or luminous infrared galaxies than in normal galaxies.
Therein lies the dilemma. If the number of unidentified sources and/or 
the number of plausible candidates is sufficiently large, what will 
constitute a sound identification? How shall we find evidence
for new populations of sources, and new members within these
populations, in the GLAST era?

\begin{figure}[ht]
   \centering
   \includegraphics[width=0.47\textwidth]{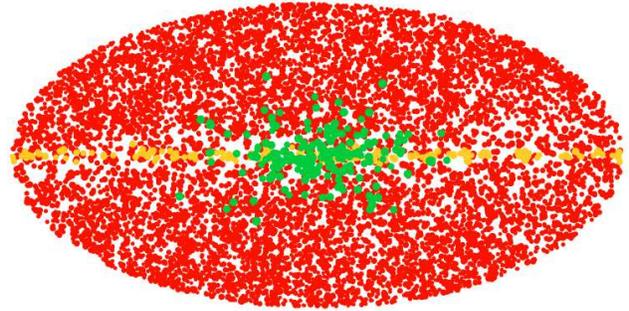} 
   \caption{Synthetic gamma-ray source catalog as in Fig.1, but symbol size has been enlarged to represent
   source location uncertainty contours as expected for a large source catalog: At almost any line of sight 
   there is a gamma-ray source found in their respective error circle. This will indicate the problem of
   probing the existence of new source populations in the GLAST-LAT era. Log~N--log~S predictions for the increased 
   instrumental sensitivity predict vastly more sources than we know of today (EGRET), and the majority of them 
   are supposedly faint sources below the EGRET detection limit, thus not tremendously better localized by GLAST-LAT.
   }
   \label{Fig.2}
\end{figure}

At present, the most successful identification scheme for $\gamma$-ray 
sources is based upon multifrequency follow-up observations, 
unless there is a given prediction of periodicity, which itself would 
unambiguously label the source if the same periodicity is found in the $\gamma$-ray 
data. The latter, however, will happen only for a fraction of GLAST-LAT detections, either 
because of the absence of contemporaneous pulsar timing solutions (in particular
for X-ray pulsars), or sufficient statistical significance for claiming periodicity 
in the still photon-limited $\gamma$-ray data, or because of shortage of precise 
theoretical predictions for testable variability patterns other than periodicity. 
Note that variability of $\gamma$-rays probes, generally, timescales, not periodicities, 
and can be used predominantly to {\it rule out} membership into classes and only
when it could be established at a significant level. For example, if a
given source {\it is} variable, we consequently assume that it is
not produced in phenomena on timescales larger than the
corresponding exposure. In essence, this will rule out all
possible counterparts producing steady $\gamma$-ray fluxes. In fact, for many 
of the theoretically anticipated LAT sources steady $\gamma$-ray emission is 
predicted. Such candidate populations are Supernova remnants (e.g., \cite{Tor03a}), 
luminous infrared galaxies (e.g., \cite{Tor04}), or galaxy clusters (e.g., \cite{Rei03}).

However, if a theoretically compatible variability time-scale exists,
it will prompt the need of carrying out follow-up observations,
which will necessarily require a considerable amount of time and
resources, without guaranteed success of achieving an unique
identification. The bottom line is that adopting this scheme, with
GLAST observations, particularly during the first year of data
taking, we may limit our capability to identify new populations of
sources if relying exclusively on multifrequency follow-up methods
for source identification.

If we have a spatial coincidence between a candidate and an
unidentified $\gamma$-ray source, and in addition there exist a
matching variability timescale between theoretical predictions for
such object and the data, then how can we, with nothing else,
definitely say that an identification was achieved? And even if we
convince ourselves to assert it, how many of such individual cases
should be found in order to claim the discovery of a new
population of sources with satisfactory statistical significance?
How would the latter be quantitatively evaluated? Not having {\it
a priori of the expected number of source detections} a criterion
by which to answer the previous questions will confront us with a
situation of ambiguity between results achieved by applying
different classification standards, with no instance to decide in
a unbiased way whether an identification has been achieved or not.

In order to overcome these predictable problems, a paradigm shift
in the way we seek the classification of $\gamma$-ray populations is
suggested. We need to define a sensitive and quantitative
criterion, by which we could identify both variable and
non-variable populations. A feasible scheme for defining such a criterion
was laid out in \cite{Tor05}, and is referred to in the following. 
Although we refer here explicitly to the case of GLAST-LAT, more particular 
$\gamma$-ray source detection with the Large Area Telescope (LAT), the scheme we 
present is adaptable to other experiments confronted with a similar combination of 
problems, for example in hemispheric neutrino astronomy.

\section{Identification of $\gamma$-ray populations}

Here we elaborate a scheme to identify and classify new
$\gamma$-ray source populations.

\subsection{What to search for?}

Starting from a given theoretical prediction of a population of
astronomical objects to be detectable above the LAT instrumental
sensitivity, we propose to impose a
\begin{itemize}
\item {\it Theoretical censorship:} we request as part of the
criterion that {\it predictions}, ideally of multiwavelength character, 
are available for a subset of the proposed class of counterparts. 
The term {\it predictions} refers here to measurable observables for 
the respective instrument.
\end{itemize}
This request is made to avoid the blind testing of populations
that may or may not produce $\gamma$-rays, but for which no other
than a spatial correlation result can be achieved a posteriori. If
there is no convincing theoretical support that a population can
emit $\gamma$-rays before conducting the search, such population
may not be sought this way. Although obvious, it should be
explicitly stated that we will not, by applying this method,
disallow the possibility of making serendipity discoveries.
Imposing of a theoretical censorship is not just a matter of
theoretical purity, but rather it is statistically motivated, as
we explain below. Such censorship applies similarly to all {\it a
priori} selection of subclasses, i.e., the imposing of cuts in
samples that are aimed to isolate the members from which we
preferably expect detectable $\gamma$-ray emission. 

\subsection{Protection of discovery potential}

By probing a large number of counterparts candidates with at least
equally large number of trials with the same data set, one {\it
will} find positive correlations, at least as a result of
statistical fluctuations (also referred to as chance capitalization). 
Then, to claim significance, one would have to check if the penalties 
that must be paid for such a finding (i.e., the fact that there were a 
number of trials that led to null results) does not overcome the 
significance achieved. Needless to say, a number of possible bias are 
expected to influence the computation of the penalties. The example here 
is ultra high energy cosmic rays (UHECRs), where there are already a number 
of dubious discovery claims from correlation studies, even when the sample
of events is small (see, e.g. \cite{Eva03}, \cite{Tor03b}). 
GLAST-LAT, and in general $\gamma$-ray astronomy, can
prepare to address this difficulty before entering the new era of
source wealth, as UHECR physics does before unblinding data from
the Pierre Auger observatory (\cite{Cla03}). In
this sense, this part of our criterion is rather similarly
defined. We require an
\begin{itemize}
\item {\it A priori protocol:} The populations that are to be
tested in the GLAST-LAT data shall be defined before the initial
data release.
\end{itemize}

A {\it protocol} is technically a budget for testing correlations. 
Every test will consume part of this budget up to a
point that, if we still proceed in testing, there can be no
statistical significant detection claim achieved anymore. A
protocol secures that a detection of a population can be made with
confidence in its statistical significance for a number of
interesting classes. As remarked by the \cite{Cla03}, when confronted 
with claims made in the absence of an a
priori protocol, one may assume that a very large number of failed
trials were made in order to find the positive results being
reported, and thus disregard the claims altogether just by denying
statistical weight. Otherwise stated, we might be asked for proof
that the penalty for failed trials has been accounted for and is
indeed below a required statistical significance. This may turn out 
to be, either very difficult to achieve or strictly impossible
because of the possible biases in penalties definitions.

Additional exploration of the same data set for expected or
unexpected populations can (and certainly will) be made, although
if the budget is spent, without the strength of immediate
discovery potential. A positive additional search must be thought
of as a way of pointing towards new populations of sources to be
tested with additional or independent sets of data
then. Here, a source catalog based on the second year of GLAST-LAT 
observations would not be independent: it will combine already 
discovered persistent sources with newly discovered ones that were 
below the instrumental sensitivity or imposed detection threshold beforehand, 
or of transient character.

Summarizing, if using the same set of data, claiming the discovery
of one population affects the level of confidence by which one can
claim the discovery of a second. Then, suppose for definiteness
that the {\it total} budget is a chance probability equal to
${\cal B}$, e.g.,  10$^{-4}$. That is, that a claim for
population(s) discovery has to be better than one having a
probability of chance occurrence equal to ${\cal B}$, and that we
want to test $A$, $B$, $C$ \ldots classes of different sources
(say, radio galaxies, starburts galaxies, microquasars, pulsars,
AGN, for a recent overview see \cite{Rei05}). The total budget can then 
be divided into individuals, a priori, chance probabilities, $P_A$, $P_B$, etc., such that
$\sum_i P_i={\cal B}$. This implies that population $i$ will be claimed 
as detected in this framework if the a posteriori, factual, probability for 
its random correlation, $P^{\rm LAT}(i)$, is less than the a priori
assigned $P_i$ (as opposed to be less only than the larger, total
budget). The $\sum_i P_i$ also accounts for
any attempt to investigate population properties of subsamples
belonging to the same object class by invoking cuts. If too many
subsamples were investigated in order to discriminate further
among the emission characteristics in an already detected source
population, such selections are on the expense of the budget, too.
Statistically dependent test shall be avoided. A minimal set of
subsamples, imposing substantially different cuts in their
selections, is the most adequate choice to maximize the chance
for statistically-significant classifications of subsamples.  

We could go a step forward and suggest to manage the budget of 
probabilities. For some populations, e.g., those which were not 
detected in EGRET observations, we can less confidently assume that 
they will be detected, or perhaps for some others, the number of 
their members may be low enough such that a detection of only several of its
individuals would be needed to claim a large significance. In this
situation we would choose a relatively higher $P_i$, so that it
would be easier to find $P^{\rm LAT}(i) < P _i$. For others, say
AGN and pulsars, we are confident that they will be detected, and thus 
we would be less willing to spent a large fraction of the discovery
budget in them. Within the protocol, we can statistically prove
that these population appear with very high confidence by
assigning a very low $P_i$ in such a way to make harder for the
test to pass. If one or more of the tests, i.e., if for several
$i$-classes, $P^{\rm LAT}(i) < P_i$, is fulfilled, the results are
individually significant. First, because we protected our search
by the a priori establishment of the protocol (a blind test) and
second, because the overall chance probability is still less than
the total budget ${\cal B}$.

We refrain ourselves here to explicitly propose which are the
populations to be tested and how large the a priori probability
assigned to each of them as well as the exact number for the total
budget ${\cal B}$ should be. This ultimately has to be carefully
studied by the GLAST-LAT collaboration for data in the proprietary period, 
although obvious choices can be compiled and argued. 
Now we proceed towards a most delicate issue, that of the treatment of 
the statistical significance of claimed detections of source populations.

\subsection{How to search and significance assessment}

The last constituent of a methodological approach to identify new
classes of $\gamma$-ray sources is the application of a

\begin{itemize}
\item {\it Common significance assessment:} we urge that a strict
statistical evaluation is mandatory before a claim of a discovery
of a new source population can be made. An objective method is
presented in the following.
\end{itemize}

We start by assessing the number of members of the relevant
candidate class being probed, for which predictions exist, that
coincide with GLAST-LAT source detections of unidentified $\gamma$-ray
sources. Let ${\cal C}(A)$ represent this number for population
$A$.
In what follows, for the sake of simplicity, we will assume that
we deal with equally probable coincidences, when a projected
position is less distant than, say, the 95\% confidence contour.

Let ${\cal N}(A)$ be the number of known sources in the particular
candidate population $A$ under analysis and ${\cal U}$ the number
of LAT detections.
Let ${\cal P}$ be the probability that in a random direction of
the sky we find a LAT source.
The probability ${\cal P}$ should take into account instrumental
detectability issues (exposure gradients, imprecision of the
diffuse emission model, etc.) as well as, at low Galactic
latitudes, expected Galactic structures.

As an example which omits the latter complications, one may use
angular coverage (the ratio between the area covered by ${\cal U}$
sources and that of the sky region upon which these sources are
projected). In what follows, we will assume that such method is in
place for LAT and that ${\cal P}$ can be computed for a given
region of the sky. Note that to compute ${\cal P}$ we do not need
any information about the candidates, but just some sensible
extrapolation of the expected number of detections of sources that
have been already identified. The value of ${\cal P}$ is obtained
a priori of checking for any population.




Whatever the method, ${\cal P}$ is expected to be small for LAT.
To give an example, if we take just a coverage assessment at high
Galactic latitudes ($|b|>10$), and we assume that there will be a
thousand detections, and that the typical size of the error box of
LAT sources is a circle of radius 12 arcmin, then ${\cal P} \sim 3
\times 10^{-3} $. At lower latitudes, we expect ${\cal P}$ to be
between 1 to 2 orders or magnitude larger. We believe that a more
careful treatment of source number predictions and the range of
expected source location uncertainties will reduce the value of
${\cal P}$ from such simple estimations. Such low values for
${\cal P}$ make the product ${\cal P} \times {\cal N}(A)$ typical
less than 1-10, for all different candidate populations. We will
refer to this product as the {\it noise expectation}, i.e., this
is the number of coincidences which one would expect even when
there is no physical connection between the LAT detections and
population $A$.

The number of excess detections above noise will be, ${\cal
E}(A)={\cal C}(A)-{\cal P} \times {\cal N}(A)$.\footnote{
Obviously,  if the number of sources is so large that ${\cal P}
\rightarrow 1$, then ${\cal E}=0$. If instead, the number of
members in the potential counterpart class is so large that ${\cal
C}(A) \rightarrow {\cal P} {\cal N}(A)$, then ${\cal E}=0$ too. In
both cases, there is no way to distinguish whether the population
is physically associated. To simplify the treatment we consider
excesses with no overlapping, i.e., coincidences between members
of population A and LAT sources that are not co-spatial with
members of other populations. In reality, the available
$\gamma$-ray observables will allow further discrimination, either
directly by reducing overlap between members of different
populations at higher photon energies (better source localization
due to narrower instrumental psf), or when the populations under
consideration become distinguishable due to their source spectra,
and variability pattern.}
%
Two cases can be distinguished. The two largest populations of
plausible candidates (pulsars and blazars) will also present the
largest number of coincidences, since it is already proven that
they do emit high energy $\gamma$-rays above LAT sensitivity,
and the populations are sufficiently large in number. Let's assume that 
there are 2000 catalogued AGNs; with the quoted value of ${\cal P}$, all 
coincidences in excess than 6 are beyond the random expectation. 
The reality of the population in the EGRET catalog make us expect 
that ${\cal C}({\rm AGN})\gg 6$, and thus that the number of excesses 
would be equally large. In this case, we are in the domain of large 
number statistics and a probability for the number of excesses to 
occur by chance, $P^{\rm LAT}({\rm AGN})$ can be readily computed.

A different case appears when the second term in the expression
for ${\cal E}(A)$ is a small quantity. Two scenarios may be found:
if the number of coincidences for that population is large
compared with the noise, we are again in the domain of large
number statistics, as in the case of AGN or pulsars. This will
--most likely-- not happen for many (or perhaps for any) of the
new populations we would like to test. Thus, in general we are in
the realm of small number statistics: we should test the null
hypothesis for a new source population against a reduced random
noise (see \cite{Fel98}, also \cite{Geh86}).\footnote{If a precise number 
of detectable sources is predicted, generally one could test the hypothesis of their
presence in the LAT catalog directly, using small number
statistics described in more detail below. However, this will unlikely
constitute the standard scenario since we will not know precisely from
theoretical arguments how many, say, of the X-ray binaries, should indeed 
be detectable. Modeling is in general not applied with an equal level of 
detail to a sufficiently high number of members in a candidate population.}

Let us analyze now an explicit example. We are testing a null
hypothesis (e.g., X-ray binaries are not LAT sources). That is
represented by 0 predicted signal events (coincidences), i.e.
total number of events equal to the background in Table 2-9 (see
leftmost columns) of \cite{Fel98}. Suppose for
definiteness that ${\cal P} \sim 3 \times 10^{-3}$ and ${\cal
N}(A)$ is equal to, say, 200, then the number of chance
coincidences (the noise or background) is 0.5. Thus, if we find
more than 5 individual members of this class (e.g. superseding the
confidence interval 0.00-4.64) correlated with LAT sources, we
have proven that the null hypothesis is ruled out at the the 95\%
CL.

Using the small number statistics formalism, we can convert the
level of confidence achieved for each population into the factual
probability, i.e., $P^{\rm LAT}({\rm X-ray\,bin})$. Subsequently,
by compfaring with the a priori budgeted requirement (i.e., is
$P^{\rm LAT}({\rm X-ray\, bin.}) < P_{\rm X-ray\, bin.}$?, we will
be able to tell whether the population has been discovered.
Clearly, if instead we find no more than 5 individual sources in
the same example, then we have no evidence by which to claim the
existence of this population at that level of
confidence.

Managing $P_A$ is equivalent to requesting different populations
to appear with different, intelligently selected, levels of
confidence. By using this method, detecting just a few members of
each class may allow to achieve significant levels of confidence,
justified by the existence of the imposed theoretical censorship
and protected by an a priori protocol. Note that at this stage
there is no variability analysis involved. If we were to add the
search on compatible variability timescales, the confidence level
of the detections will even improve.

\section{Concluding remarks}

The proposed criterion for identification of $\gamma$-ray source
populations integrates three different parts:
1) A {\it theoretical censorship} that prohibits executing
repeated searches that would reduce the statistical significance
of any possible positive class correlation.
2) An {\it a priori protocol} that protects the significance by
which to claim the discovery of a number of important population
candidates and gives guidelines as to how to manage the
probability budget
3) A {\it significance assessment} that assigns probabilities both
in the large and in the small numbers statistical regime.

It is useful to note that LAT will be in a privileged position to
actually identify new population of sources. If LAT would have an
additional order of magnitude better sensitivity, without
significant improvement in angular resolution, a situation similar
to the GRB case would appear, i.e., a flat distribution of
unidentified sources with a few privileged individuals only which
are extensively studied in multifrequency studies. Essentially, we
would find a $\gamma$-ray source coinciding with the position of
every member of any population under consideration. And thus, we
would lack the capability to achieve discoveries by correlation
analysis. This is, perhaps, already indicating that a next
generation high energy $\gamma$-ray mission after GLAST-LAT might
not be exclusively sensitivity-driven if no significant
improvement in angular resolution can be achieved.

The potential of this methodological procedure is not limited to
the anticipated cases explicitly discussed here. By applying the
proposed scheme, one can also check spurious classifications in an
objective way, and test subsamples among the expected classes of
sources (e.g., FSRQs in correspondence of their peak radio flux,
or BL Lacs in correspondence of their peak synchrotron energy,
i.e. LBLs vs. HBLs, galaxy clusters in correspondence of their
X-ray brightness). Summarizing, the portrayed identification
scheme is not exclusively elaborated for source populations in
high-energy $\gamma$-rays. It's a methodological approach to be
generally applicable if the identification of source populations
among a complex astrophysical dataset can only be achieved by a
statistically sound discrimination between candidate classes.

\label{discussion}
\section*{Acknowledgments}
DFT has been supported by Ministerio de Educaci\'on y Ciencia (Spain) 
under grant AYA-2006-00530, as well as by the Guggenheim Foundation.

\end{document}